\documentstyle[12pt,epsf,axodraw]{article}

\newcommand{\z}{&\hspace*{-8pt}}

\begin{document}

\begin{flushright}
DESY 99-008\\
BI-TP-99/07 \\
\end{flushright}

\begin{center}


{\Large \bf Non factorizable $O(\alpha\alpha_s)$ corrections
 to the process $Z\to b\bar{b}$}

\vskip 10mm

J.~Fleischer\footnote{~E-mail: fleischer@physik.uni-bielefeld.de}$\,{}^a$,
F.~Jegerlehner\footnote{~E-mail: fjeger@ifh.de}$\,{}^b$,
M.~Tentyukov\footnote{~E-mail: tentukov@physik.uni-bielefeld.de}%
\footnotemark[5]$\,{}^a$,
O.~L.~Veretin\footnote{~E-mail: veretin@ifh.de}%
\footnote{~Supported by BMBF under 05~7BI92P(9)}$\,{}^{ab}$

\vskip 10mm

${}^a${\it ~Fakult\"at f\"ur Physik, Universit\"at Bielefeld,
D-33615 Bielefeld, Germany.}\\[5mm]
${}^b${\it DESY Zeuthen, Platanenallee 6, D-15738 Zeuthen, Germany.}\\[30mm]

\begin{abstract}

  We evaluate non factorizable $O(\alpha\alpha_s)$ corrections
to the process $Z\to b\bar{b}$ due to the virtual t-quark.
All two-loop vertex diagrams with $W$'s and charged ghosts $\Phi$'s
are included. They are evaluated in the large top-mass expansion
up to the $10^{\rm th}$ order. Gluon
Bremsstrahlung is taken into account by integrating over the whole
phase space. All calculations, including Bremsstrahlung, are done in
dimensional regularization. The expansion coefficients of the
large mass expansion are given in closed form. Their expansion
in $y=m_Z^2/4m_W^2$ is in agreement with the coefficients up to 
$O(m_W^6/m_t^6)$ as given by Harlander et al. \cite{Zbb1}.
\end{abstract}

\end{center}



\thispagestyle{empty}
\setcounter{page}0
\newpage


The large statistics of the LEP I experiments yielded precise values
for the partial $Z$-decay width into $b$-quarks; expressed in terms of
the ratio $R_b = 0.21656 \pm 0.00074$ \cite{Karlen} this is a
precision of $\sim 0.3 \%$. Therefore precise high order calculations
in the Standard Model (SM) are needed.

  The partial width of the $Z$-boson decay into a
quark-antiquark pair can be parametrized as
\begin{eqnarray}
\label{partialGamma}
\Gamma = 
\tilde\Gamma 
  \left( 1 + \frac{\alpha}{\pi}\delta_{\rm EW} + \dots \right)
  \left( 1 + \frac{\alpha_s}{\pi} + \dots \right)
  \left( 1 + \frac{\alpha}{\pi}\frac{\alpha_s}{\pi}
          \Delta \right),
\end{eqnarray}
where 
\begin{equation}
\label{tildeGamma}
 \tilde\Gamma = \frac{\alpha N_c m_Z}{12s^2c^2}(v^2+a^2)
     \Bigl(1+{\Pi'}_Z(m_Z^2)\Bigr)^{-1}.
\end{equation}
  (\ref{partialGamma}) defines the non factorizable contribution $\Delta$.
$s,c$ are the sinus and cosinus of the weak mixing angle;
$N_c=3$ is the color factor;
$v$ and $a$ are vector and axial couplings
related to isospin $I_3$ and charge $Q$ of the fermion by
\begin{equation}
 v=I_3 - 2Qs^2,\qquad a=I_3.
\end{equation}
The renormalized self-energy $\Pi_Z$ in (\ref{tildeGamma})
accounts for the universal correction due to $Z$-boson renormalization.

   The electroweak correction $\delta_{\rm EW}$ has been calculated
long ago \cite{EW0} (see also \cite{EW1}). 
The QCD correction factor $(1+\frac{\alpha_s}{\pi}+\dots)$
is known now up to $O(\alpha_s^3)$ in the massless limit
\cite{QCD0} and also at $O(m_q^2/s)$ \cite{QCD1}. However, electroweak and
QCD corrections do not factorize exactly. Therefore the correction
factor $\Delta$ appears. This correction for the quarks of
u-,d-,c- and s-type was found in \cite{light1}, where the masses
of quarks were neglected compared to $m_Z$. For the b-quark case
it is necessary to consider diagrams with the virtual t-quark.
The leading term proportional to $m_t^2$ has been calculated
in \cite{leading} and the term proportional to $\log m_t^2$ in \cite{logmt}.
See also recent work \cite{3loop} where leading $m_t$ corrections
$O(m_t^2 G_F \alpha_s^2)$ were calculated for $Z\to b\bar{b}$ decay mode.

  In the following we are interested only in contributions
with the top quark. However, this alone is neither finite nor
gauge invariant (see e.g. the discussion in \cite{Zbb1,light1}). 
Instead we have to consider all diagrams with $W$-boson exchange. 
In other words we have the gauge invariant decomposition
\begin{equation}
  \Delta = \Delta^Z + \Delta^W,
\end{equation}
where $\Delta^{Z,W}$ denotes contributions due to $Z,W$-boson exchange,
respectively. 

  Recently in \cite{Zbb1} the subleading terms for $\Delta_b$
were found. The expansion in $1/m_t^2$ has been made up to
$(1/m_t^2)^3$. Each coefficient of the expansion in turn
is expanded in the small parameter $y=m_Z^2/4m_W^2$ by a large
mass expansion of subdiagrams. The calculations
of \cite{Zbb1} were done by cutting the 3-loop two point function
of the $Z$-boson. 

  In a sense our approach is complementary to
that of \cite{Zbb1}: we calculate directly the 
amplitude of the $Z\to b \bar{b}$
process and integrate over the final state phase space. In addition
we have to add the gluon Bremsstrahlung to form an IR finite quantity.
We also use the standard large mass expansion (LME) technique \cite{LME}
for both real and virtual gluon processes: $1/m_t^2$ is considered
as small parameter and the expansion in $1/m_t^2$ has
been performed up to $(1/m_t^2)^{10}$. We do not, however, perform a further
expansion of subdiagrams, therefore we obtain closed expressions
for the expansion coefficients.

  However, after reexpansion we fully agree on the first few
coefficients given in \cite{Zbb1}. We also agree on the numerical
estimates following from \cite{Zbb1} and from our work.
A detailed comparison was given in \cite{Barselona}. 
However in the present work the results are given in the parametrization 
(\ref{partialGamma}) which seems to be very natural.

  In a recent paper \cite{FKV} some scalar 
2-loop vertex diagrams relevant for 
$\Delta_b$ were analysed. There it was found that the series in $1/m_t^2$
has a rather bad behaviour (expansion coefficients grow like $4^n$).
Therefore one would assume that higher terms of the expansion are needed.
However, for the complete 
physical quantity this is not the case. The 'dangerous'
terms cancel as well as the auxiliary structures 
like the polylogarithms ${\rm Li}_2,{\rm Li}_3$
which enter separate diagrams but not the sum of all contributions.

  In the course of our calculation we used a 
program written in FORM \cite{FORM}.
The input for the FORM procedures was generated by the 
C program DIANA \cite{DIANA}.

  The amplitude of the process
$Z(q)\to \bar{b}(p_1) + b(p_2)$ is given by
\begin{equation}
  {\cal M} = \epsilon^\mu(q)\,T_\mu(q,p_1,p_2),
\end{equation}
where $\epsilon^\mu(q)$ is the $Z$-boson wave function  and the
amplitude $T_\mu$ reads
\begin{eqnarray}
\label{Tmu}
   T_\mu = -i\frac{e}{2sc} \bar{b}(p_1) \Bigl[
   \gamma_\mu\, v(q^2) - \gamma_\mu\gamma_5\, a(q^2) \Bigr] b(p_2)
\end{eqnarray}
where $e$ is the electric charge; $\bar{b},b$ are wave functions of
b-quarks. At the tree level the couplings $v_b$ and $a_b$ are given by
\begin{equation}
\label{gtree}
  v_b=-\frac12+\frac23 \sin^2\Theta_W, \qquad  a_b=-\frac12,
\end{equation}
while in higher orders (\ref{gtree}) gets corrections to both
real and imaginary parts. 

  Evaluating the width from (\ref{Tmu}), using the standard rules, we get
in $d=4-2\varepsilon$ dimensional space-time
\begin{eqnarray}
\label{GammaZbb}
  \Gamma \z=\z
    \left( \frac{\bar{\mu}^2}{m_Z^2}\right)^{(4-d)/2} e^{\varepsilon\gamma_E}
    \frac{\alpha N_c m_Z}{12s^2c^2}
    \frac{\Gamma(d/2)}{\Gamma(d-2)} 
    \Bigl( |v|^2 + |a|^2 \Bigr),
\end{eqnarray}
where $\bar{\mu}$ is an arbitrary parameter with the dimension of a mass.
It is related to the parameter $\mu$ of dimensional regularization
by $\bar{\mu}^2 = 4\pi e^{-\varepsilon\gamma_E}\mu^2$.

  At 1-loop level the diagrams contributing to this process
are shown in Fig.1. They were evaluated in \cite{EW0,EW1}.
%
%
\begin{figure}[h]
\centerline{\vbox{\epsfysize=50mm \epsfbox{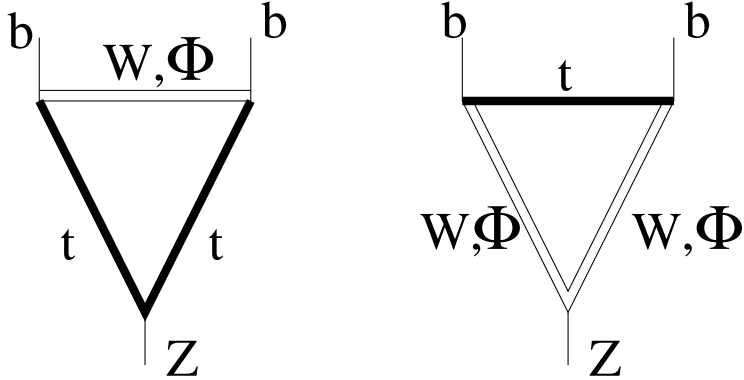}}}
\vspace*{3mm}
\noindent
\caption{\label{diagrams} 1-loop diagrams with the t-quark contributing
to the $Zb\bar{b}$ process of order $O(\alpha)$. Diagrams of
order $O(\alpha\alpha_s)$ are obtained by adding gluon lines.
        }
\end{figure}

  Mixed $O(\alpha\alpha_s)$ corrections are obtained from
the diagrams of Fig.1 by adding in all possible ways a gluon line.
In this way we obtain 2-loop vertex diagrams as well as 
1-loop Bremstrahlung diagrams.

  To obtain the renormalized expressions, for our
purpose it is enough to renormalize the mass 
of the t-quark to order $O(\alpha_s)$ and the
wave function of the b-quark up to $O(\alpha\alpha_s)$.

  The renormalization constant of $m_t$ in the on-shell scheme
is given by
\begin{equation}
\label{trenormalization}
Z_{m_t} = 1-\frac{\alpha_s}{\pi}C_F \Biggl(
   \frac{3}{4\varepsilon} + 1 - \frac34 \log\frac{m_t^2}{\bar{\mu}^2} \Biggr). 
\end{equation}

  The effect of the b-quark renormalization is effectively
performed as
\begin{eqnarray}
\label{brenormalization}
  (v-a)_R = Z_{\rm b}(v-a)_0,
\end{eqnarray} 
with the $b$-quark renormalization constant $Z_b$
(see appendix),
while the right handed combination $(v+a)$ remains
unrenormalized. 

  To make the width IR finite, in addition 
to the nonradiating process we have to add the
soft contribution from the Bremsstrahlung process
\begin{equation}
\label{Zbbg}
   Z \longrightarrow b+\bar{b}+g.
\end{equation}

  Actually we will include the hard gluon emittion as well by
integrating over the full gluon phase space. 
  Of particular interest is the calculation of the Bremsstrahlung
  in terms of the LME.
  The kinematics of the (gluon-) Bremsstrahlung for the process
(\ref{Zbbg})
  is given by

\begin{equation}
  q\to p_1+p_2+p_3,\qquad  p_1^2=p_2^2=p_3^2=0.
\end{equation}

  Thus we have 3 invariants
\begin{equation}
  p_1p_2,\qquad p_2p_3,\qquad p_1p_3.
\end{equation}

  We are interested in the integrated Bremsstrahlung.
The corresponding width in $d$ dimensions can be
written as
\begin{eqnarray}
\label{GammaBrem}
 \Gamma_{\rm Br} \z=\z \left( \frac{\bar{\mu}^2}{m_Z^2}\right)^{4-d}
    e^{2\varepsilon\gamma_E}
    \frac{m_Z}{768\pi^3} \frac{1}{\Gamma(d-2)}  \nonumber\\
\z\z \int_0^1 dx\,dy\,
    x^{d-3} (1-x)^{d/2-2} y^{d/2-2} (1-y)^{d/2-2}
    \left| M_{\rm Br}\right|^2,
\end{eqnarray}
where $M_{\rm Br}$ is the Bremsstrahlung amplitude
and the invariants can be expressed in terms of $x$ and $y$ as
\begin{eqnarray}
    p_1p_2 &=& \frac{m_Z^2}{2}x(1-y)\,,  \nonumber\\
    p_2p_3 &=& \frac{m_Z^2}{2}(1-x)\,,   \nonumber\\
    p_1p_3 &=& \frac{m_Z^2}{2}xy   \,.   \nonumber
\end{eqnarray}

  It should be noted that neither (\ref{GammaZbb}) nor (\ref{GammaBrem})
are finite separately after the renormalization. They both have
IR divergences up to $O(1/\varepsilon^2)$. Therefore it is quite
important to use general $d$-dimensional expressions and take
the limit $d\to4$ only after adding up (\ref{GammaZbb}) and (\ref{GammaBrem}).
Otherwise some finite contributions will be lost.

  We have checked by  explicit calculation that indeed the result is
finite in the sum of $\Gamma$ and $\Gamma_{\rm Br}$ and our result for
$\Delta_b^W$ reads
\begin{eqnarray}
\z\z \Delta_b^W = \frac{C_F}{(v_b^2+a_b^2)s^2}\Biggl[ \nonumber\\
\z\z
    \frac{m_t^2}{m_W^2} \Biggl\{
        \zeta_2 (\frac{1}{16}+\frac{1}{32y}) 
             \Biggr\}  
 \nonumber\\
\z\z
       +  \Biggl\{
        - \frac{3245}{11664}y 
        - \frac{7499}{46656} 
        - \frac{1009}{93312y}  
        + \zeta_2 ( 
            \frac{53}{324}y
          + \frac{173}{1296} 
          + \frac{67}{2592y}
              )
    \nonumber\\
    \z\z\qquad
       + L_c (  
              - \frac{7}{1944}y 
              - \frac{7}{1944} 
              - \frac{7}{7776y}
             ) 
       + L_t (  
              - \frac{7}{1944}y 
              - \frac{7}{1944} 
              - \frac{7}{7776y} 
             )
             \Biggr\}  
 \nonumber\\
\z\z
    + \frac{m_W^2}{m_t^2} \Biggl\{
         - \frac{23939}{21600}y^2 
         - \frac{453539}{777600}y 
         - \frac{262937}{1555200} 
         - \frac{89}{1152y}
    \nonumber\\
    \z\z\qquad
         + \zeta_2 (
              \frac{11}{18}y^2
            + \frac{13}{144}y
            + \frac{257}{864}
            + \frac{175}{864y}
               )
    \nonumber\\
    \z\z\qquad
       + I_0  (  
               \frac{13}{216}y^2
             + \frac{13}{48}y 
             - \frac{13}{96} 
             - \frac{299}{1728y} 
             - \frac{13}{576y^2} 
              )
    \nonumber\\
    \z\z\qquad
       + L_c  (  
               \frac{17}{38880}y^2 
             - \frac{823}{38880}y
             - \frac{3343}{155520}
             - \frac{7}{1296y} 
             )
    \nonumber\\
    \z\z\qquad
       + L_t (  
               \frac{2357}{38880}y^2 
             + \frac{8537}{38880}y 
             - \frac{45823}{155520} 
             - \frac{1009}{5184y} 
             )
      \Biggr\}
 \nonumber\\
\z\z
    + \frac{m_W^4}{m_t^4} \Biggl\{
       - \frac{652232029}{178605000}y^3 
       - \frac{938540803}{158760000}y^2 
       + \frac{362957621}{317520000}y 
       + \frac{1198673}{1166400} 
       - \frac{1099}{3888y}
    \nonumber\\
    \z\z\qquad
       + \zeta_2 (
            \frac{4838}{2025}y^3 
          + \frac{4769}{1080}y^2
          - \frac{14539}{10800}y
          - \frac{239}{324}
          + \frac{10}{27y}
             )
    \nonumber\\
    \z\z\qquad
       + I_0 ( 
          - \frac{91}{540}y^3 
          - \frac{383}{540}y^2 
          + \frac{4139}{4320}y 
          + \frac{1021}{4320} 
          - \frac{533}{1920y} 
          - \frac{5}{128y^2} 
             )
    \nonumber\\
    \z\z\qquad
       + L_c L_t ( \frac{1}{36}y + \frac{1}{36}+ \frac{1}{144y} )
       + L_t^2      ( \frac{1}{36}y + \frac{1}{36}+ \frac{1}{144y} )
    \nonumber\\
    \z\z\qquad
       + L_c (  
          - \frac{31}{283500}y^3 
          + \frac{283}{212625}y^2 
          - \frac{105443}{3402000}y 
          - \frac{623}{19440} 
          - \frac{7}{864y} 
            )
    \nonumber\\
    \z\z\qquad
       + L_t (  
          - \frac{23903}{141750}y^3
          - \frac{1060861}{1701000}y^2 
          + \frac{3447757}{3402000}y 
          - \frac{11551}{38880} 
          - \frac{811}{1728y} 
            )
          \Biggl\}
    \nonumber\\
    \z\z
        + O\left(\frac{m_t^6}{m_W^6}\right) \Biggr]
\end{eqnarray}
where $L_t=\log(m_t^2/m_W^2)$, $L_c=\log c^2$
and $y=m_Z^2/4m_W^2$. The only function that enters the answer
is
\begin{equation}
  I_0 = -\frac12\int_0^1 \frac{\log(1-t y)}{\sqrt{1-t}}
       \,dt.
\end{equation}
We do not give here higher coefficients in analytic form
because of their complexity. Instead we give below 10 coefficients
of the $1/m_t^2$ expansion numerically.
For $m_t=175$GeV, $m_W=80.33$GeV and $m_Z=91.187$GeV we obtain
\begin{eqnarray}
\Delta_b^W \z=\z 
    \frac{4.1878}{t}
   + 2.3057
   - 8.0270 t
   - 28.0471 t^2 \nonumber\\
\z-\z 
     39.5864 t^3
   - 32.7842 t^4
   - 8.7501 t^5
   + 20.0429 t^6 \nonumber\\
\z+\z
     36.7551 t^7
   + 37.3991 t^8
   + 65.1923 t^9
   + 307.874 t^{10}\,, 
\end{eqnarray}
where $t=m_W^2/m_t^2\sim 0.21$.

  Separating the leading term from the higher order ones, 
we may write this as
\begin{equation}
\label{DeltaWb}
\Delta_b^W = 19.8764-1.0654 = 18.8109,
\end{equation}
which tells us that
the terms of higher order amount to only $\sim 5\%$ of the
leading term. Obviously also the series in $1/m_t^2$ converges quite
rapidly. The reason, however, for the smallness of the correction is
the alternation in sign of the various higher order contributions.



For completeness and for comparison we also present the numbers for
the non factorizable contributions of the lighter quarks and the
Z-contribution of the b-quark. The result for these
contributions are taken from \cite{light1}.
To slightly improve them, we performed a Pad\'{e} approximation for
the expansions in $x=1$ for $\Delta^Z$ and $x=m_Z^2/m_W^2$
for $\Delta^W$ (however this gives only minor changes of order
of few percents in formulae (\ref{DeltaZlight}) and (\ref{DeltaWlight}) ). 
For $\Delta^Z$ we have
\begin{eqnarray}
\label{DeltaZlight}
 \Delta^Z = \left\{ 
      \begin{array}{ll}
        -0.489 & \mbox{for u,c} \\
        -0.796 & \mbox{for d,s,b} 
       \end{array}
             \right. 
\end{eqnarray}
  
  Taking into account (\ref{DeltaWb}), we have for $\Delta^W$ 
\begin{eqnarray}
\label{DeltaWlight}
 \Delta^W = \left\{ 
      \begin{array}{ll}
        -3.652 & \mbox{for u,c} \\
        -3.745 & \mbox{for d,s} \\
        +18.811 & \mbox{for b} 
       \end{array}
             \right. 
\end{eqnarray}
The above numbers demonstrate again that due to the heavy
top quark the $\Delta_b^W$ contribution is larger by almost an
order of magnitude than all the other contributions. The smallness of
the subleading terms in $1/m_t^2$ compared is 
surprising, however. This is a highly nontrivial result. It is, e.g.,
in contrast to the results found in Ref. \cite{Degrassi}, where an 
$O({\alpha}^2 m_t^2/m_W^2)$ calculation of $\Delta r$ was performed
and it was found that this correction is of the same order as the leading
$O({\alpha}^2 m_t^4/m_W^4)$ correction.

  A detailed comparison of our results with those of Ref. \cite{Zbb1}
is given in \cite{Barselona}.


\appendix

\section{Renormalization of the b-quark wave function}

  In this section we evaluate the virtual top quark 
contribution to the wave function renormalization
constant $Z_{\rm b}$ for the b-quark field
in the on-shell scheme. We need this constant up to
order $O(\alpha\alpha_s)$. At 1-loop order two diagrams contribute.
They are shown in Fig.2. Adding in all possible ways one gluon 
line we get 8 diagrams of order $O(\alpha\alpha_s)$. 
%
%
\begin{figure}[h]
\centerline{\vbox{\epsfysize=30mm \epsfbox{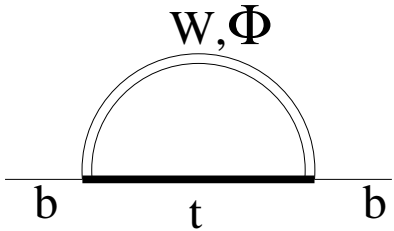}}}
\vspace*{3mm}
\noindent
\caption{\label{diagrams} 1-loop diagrams with the $t$-quark 
contributing to the b-quark wave function
renormalization in $O(\alpha)$. Diagrams in
$O(\alpha\alpha_s)$ are obtained by adding one gluon line.
        }
\end{figure}

  The self-energy of the b-quark reads
\begin{equation}
  \hat{\Sigma}(p) = \hat{p}\,\frac{1-\gamma_5}{2}\,\Sigma_L(p)
                   +\hat{p}\,\frac{1+\gamma_5}{2}\,\Sigma_R(p)\,.
\end{equation}
By explicit calculation we find $\Sigma_R=0$. Therefore 
the renormalization constant $Z_{\rm b}$ in the on-shell scheme 
($m_{\rm b}=0$) reads
\begin{eqnarray}
  Z_{\rm b} = 1+ \frac{1-\gamma_5}{2}\,\Sigma_L(0) 
   = 1+ \frac{1-\gamma_5}{2} \frac{\alpha}{4\pi s^2} 
     \Biggl( z_1 + \frac{\alpha_s}{4\pi} C_F\, z_2 \Biggr).
\end{eqnarray} 

  For the sake of completeness we give below the 
first 4 coefficients of the expansion of $z_{1,2}$ in $m_W^2/m_t^2$.
Note that we have to keep terms up to $O(\varepsilon)$
in the 1-loop part and terms up to $O(1)$ in the 2-loop part.
\begin{eqnarray}
\z\z z_{1} = \nonumber\\
\z\z
         \frac{m_t^2}{m_W^2} \Biggl\{
          - \frac{1}{4\varepsilon}
          + \frac14 L_\mu
          + \frac14 L_t
          - \frac38 \nonumber\\
    \z\z\qquad
          + \varepsilon\Bigl( 
             - \frac18 \zeta_2
             - \frac14 L_t L_\mu
             + \frac38 L_t
             - \frac18 L_t^2
             + \frac38 L_\mu
             - \frac18 L_\mu^2
             - \frac{7}{16} \Bigr)
             \Biggr\}  
 \nonumber\\
\z\z
       +  \Biggl\{
          - \frac{1}{2\varepsilon}
          + L_t
          + \frac12 L_\mu
          - \frac12  \nonumber\\
  \z\z\qquad
          + \varepsilon\Bigl(
             - \frac14 \zeta_2
             - L_t L_\mu
             + L_t
             - \frac12 L_t^2
             + \frac12 L_\mu
             - \frac14 L_\mu^2
             - \frac12
                   \Bigr)
         \Biggr\}
  \nonumber\\
\z\z
       + \frac{m_W^2}{m_t^2} \Biggl\{
            \frac74 L_t
          - \frac34
          + \varepsilon\Bigl(
             - \frac74 L_t L_\mu
             + \frac{13}{8} L_t
             - \frac78 L_t^2
             + \frac34 L_\mu
             - \frac58
                       \Bigr)
             \Biggr\}
  \nonumber\\
\z\z
       + \frac{m_W^4}{m_t^4} \Biggr\{
            \frac52 L_t
          - \frac34
          + \varepsilon\Bigr(
             - \frac52 L_t L_\mu
             + \frac94 L_t
             - \frac54 L_t^2
             + \frac34 L_\mu
             - \frac58 
                      \Bigl)
            \Biggr\}
  \nonumber\\
\z\z     + O(\frac{m_W^6}{m_t^6}), \nonumber\\
\z\z \nonumber\\
\z\z z_{2} = \nonumber\\
\z\z
      \frac{m_t^2}{m_W^2}\Biggl\{
            \frac{3}{4\varepsilon^2}
          + \frac{1}{\varepsilon} \Bigl(
             - \frac32 L_t
             - \frac32 L_\mu
             + \frac52
                    \Bigr)  \nonumber\\
  \z\z\qquad
          + 4
          + \frac34 \zeta_2
          + 3 L_t L_\mu
          - 5 L_t
          + \frac32 L_t^2
          - 5 L_\mu
          + \frac32 L_\mu^2
                \Biggr\}
  \nonumber\\
\z\z
      + \Biggl\{
            \frac{3}{4\varepsilon}
          - \frac18
          - 6 \zeta_2
          - \frac32 L_t
          - \frac32 L_\mu
          \Biggr\}
%
     + \frac{m_W^2}{m_t^2}\Biggl\{
            \frac{27}{4}
          - \frac{21}{2} \zeta_2
          + \frac{33}{4} L_t
          \Biggr\}
  \nonumber\\
\z\z
      + \frac{m_W^4}{m_t^4}\Biggl\{
            \frac{51}{4}
          - 15 \zeta_2
          + \frac{39}{2} L_t
                  \Biggr\}
    + O(\frac{m_W^6}{m_t^6}),
\end{eqnarray}
where $L_t=\log(m_t^2/m_W^2)$ and $L_\mu=\log(\bar{\mu}^2/m_W^2)$.

\end{document}